\documentclass[prb,twocolumn,showpacs,floatfix,amsmath,amssymb,superscriptaddress]{revtex4}
\usepackage{latexsym}
\usepackage{graphicx}
\usepackage{tabularx}
\usepackage{times}
\usepackage{color}
\usepackage{amsmath}
\usepackage{dcolumn}
\usepackage{latexsym,amsmath,amssymb,bm,euscript}
\newcommand{\abs}[1]{\vert #1 \vert}
\newcommand{\bra}[1]{\langle #1 \vert}
\newcommand{\ket}[1]{\vert #1 \rangle}
\newcommand{\bracket}[2]{\langle #1 \vert #2\rangle}

\newcommand{\expt}[1]{\langle #1 \rangle}

\newcommand{\shortdot}{\!\cdot\!}
\begin{document}

\title{Chern-Simons-Higgs transitions out of Topological Superconducting Phases}

\author{David J. Clarke}
\affiliation{Microsoft Research, Station Q, University of California, Santa Barbara, CA 93106, USA}
\affiliation{Condensed Matter Theory Center, Department of Physics, University of Maryland, College Park, Maryland 20742, USA}
\author{Chetan Nayak}
\affiliation{Microsoft Research, Station Q, University of California, Santa Barbara, CA 93106, USA}
\affiliation{Department of Physics, University of California, Santa Barbara, CA 93106, USA}
\begin{abstract}
In this study, we examine effective field theories of superconducting phases with topological order, making connection to proposed realizations of exotic topological phases
(including those hosting Ising and Fibonacci anyons) in superconductor-quantum Hall heterostructures. Our effective field theories for the non-Abelian superconducting states are non-Abelian Chern-Simons theories
in which the condensation of vortex-quasiparticle composites lead to the associated Abelian quantum Hall states.
This Chern-Simons-Higgs condensation process is dual to the emergence of superconducting non-Abelian topological phases in coupled chain constructions.
In such transitions, the chiral central charge of the system generally changes, so they fall outside the description of bosonic condensation transitions
put forth by Bais and Slingerland \cite{Bais09}
(though the two approaches agree when the described transitions coincide).
Our condensation process may be generalized to Chern-Simons theories based on arbitrary Lie groups, always describing a transition from a
Lie Algebra to its Cartan subalgebra. We include several instructive examples of such transitions.
\end{abstract}
\maketitle

\section{Introduction}

Some phases of two-dimensional electron systems support Fibonacci anyons, which are quasiparticles with the property that a collection of $N$ Fibonacci anyons at fixed positions has a Hilbert space of degenerate states of dimension given by the $N^{\rm th}$ Fibonacci number. When the two Fibonacci anyons are exchanged in a counter-clockwise manner, the wavefunction of the system changes by a phase $e^{4\pi i/5}$ if they fuse to form a topologically-trivial excitation (i.e. one that can be created by a local operator) or $e^{-3\pi i/5}$ if they fuse to form a Fibonacci anyon. As a result, quantum information can be stored in a collection of Fibonacci anyons and braiding supplies a universal gate set for quantum computation.

The computational universality of braiding operations endows Fibonacci anyons with a significant advantage over Ising anyons -- quasiparticles supporting Majorana zero modes. In the latter case, gates not provided by braiding (such as a $\pi/8$ phase gate) are needed for computational universality, and the purification of these noisy gates entails substantial computational overhead. However, Ising anyons have the advantage that they can be realized in free fermion models, which has led to the proliferation of candidate systems in which the existence of Majorana zero modes can be reduced to determining the topological properties of a free fermion band structure\cite{Alicea12a,Beenakker2013a,DasSarma15}. This is not the case for Fibonacci anyons. However, recent progress \cite{Teo14,Mong14a} on ``coupled chain'' constructions has facilitated controlled calculations predicting the existence of Fibonacci anyons in superconductor-fractional quantum Hall hybrid systems. The tractability of one-dimensional conformal field theories is bootstrapped to analyze a two-dimensional topological phase.

In this paper, we give an intrinsically two-dimensional formulation of such systems. We construct a non-Abelian Chern-Simons-Higgs effective field theory for the transitions from a Fibonacci superconducting phase to an Abelian quantum Hall phase. This theory also encompasses transitions to other nearby phases. Our effective theory generalizes Abelian Chern-Simons-Higgs theories of the hierarchy of fractional quantum Hall states \cite{Haldane83,Halperin84,Jain07} (see also Ref. \onlinecite{Bonderson08a}
for a non-Abelian generalization). The central charge generally changes in the transitions in our theory but, in those special cases in which it doesn't, our results agree with those of Slingerland and Bais.\cite{Bais09}

Our starting point is a Chern-Simons theory for a superconductor in which the symmetries of the Nambu spinor become emergent gauge symmetries at low energies. As a consequence, the edge of the system has gapless excitations with spin and particle-hole symmetries. We find that the Fibonacci superconductor constructed in Ref. \onlinecite{Mong14a} is among these phases. We consider possible condensates that drive the system from this phase. The resulting phases are generally quantum Hall phases which can be Abelian or non-Abelian. In the latter case, the states can
be understood as examples of the two-component non-Abelian quantum Hall states of Ref. \onlinecite{Vaezi14}. Thus, this theory gives a dual description to the transition from an Abelian quantum Hall state to a Fibonacci superconductor \cite{Mong14a}. We briefly discuss the generalization to other gauge groups, which includes, as a special case, a pure Fibonacci theory.

\section{Coupled chain constructions and the Bulk-Edge Correspondence}\label{Sec-CC}

We begin by briefly reviewing the superconducting Fibonacci phase that is a principal motivation for this paper. This is a superconducting phase that coexists with topological order. The topological order has a single non-trivial quasiparticle type, a Fibonacci anyon. Although the ultimate long-wavelength physics is quite simple, it is built on a more complicated intermediate scale scaffolding that is revealed in the richness of
nearby phases. Indeed the only known microscopic route to this phase involves both superconducting and Abelian fractional quantum Hall physics, as we now briefly review. Our ultimate construction will be, at least conceptually, dual to this.

Following the pioneering work of Teo and Kane \cite{Teo14}, it has been established that one may analyze 2D topological phases from the perspective of coupled critical 1D
chains \cite{Mong14a,Sagi14,Stoudenmire15,Mross15}.
These coupled chain constructions proceed by considering a series of wires each tuned
to the critical point of a 1+1-D conformal
field theory (CFT). Interactions between the wires serve to couple the left movers of one chain to the right movers of the next, leading to a gapped bulk surrounded by a gapless edge which is a chiral version of the same CFT. Crucially, the inter-wire couplings can only transfer quasiparticles of a given type between wires if those quasiparticles can exist in the medium in which the wires are embedded.

For instance, in the coupled wire construction of the superconducting Fibonacci (SFib) phase described in Ref.~\onlinecite{Mong14a}, the CFT is the $\mathbb{Z}_3$-parafermion CFT with central charge $c=4/5$, and the couplings hop $\mathbb{Z}_3$ parafermions (which are Abelian anyons) from one chain to the next. Such couplings would not be allowed if the medium between the wires were vacuum, since the vacuum does not support $\mathbb{Z}_3$ parafermionic excitations. For this reason, the proposal of Ref.~\onlinecite{Mong14a} could not be carried out using local Potts spins, despite the fact that the critical point of the $\mathbb{Z}_3$ Potts model is described by the appropriate CFT. Rather, the wires must be constructed within a parent Abelian phase such as the $\nu=2/3$ FQH state that supports the $\mathbb{Z}_3$ part of the theory.

Thus, the superconducting Fibonacci phase can be viewed as a descendant of a parent $\nu=2/3$ FQH state and, as a result, there are actually two edges between it and the trivial vacuum. The inner edge is given by the CFT of the wire construction, and the outer edge is given by the parent theory. In the case of the superconducting Fibonacci phase, the inner edge is given by the $c=4/5$ $\mathbb{Z}_3$-parafermion chiral CFT while the outer edge is given by the corresponding chiral CFT of the parent $\nu=2/3$ FQH state. The simplest theory allowing the necessary interwire interactions
is a bosonic $(2,2,1)$ fractional quantum Hall state with central charge $c=2$. The full compound edge between the SFib phase and the vacuum is (up to edge reconstruction) a pure Fibonacci CFT ($\mathrm{Fib}$) at central charge $14/5$. Such a theory may be described in the bulk by a $G_2$ level $1$ Chern-Simons theory, which we shall discuss in greater detail in Sec.~\ref{secG2}.

For now, we focus our interest on electronic, rather than bosonic, parent theories. The simplest fermionic theory that can be a parent state of the SFib phase is the $(1,1,2)$ FQH state at $\nu=2/3$. This state contains an unpolarized electronic excitation in addition to the $\mathbb{Z}_3$ Abelian excitations necessary to host the coupled wire construction of the SFib states. On the edge, the theory may be factored into the product of an Abelian $\mathbb{Z}_3$ theory and a theory containing only an electron (see Appendix~\ref{appfermions}). The combined theory must have chiral central charge
$c-\overline{c}=0$; since the Abelian $\mathbb{Z}_3$ has $c=2$, the ``electron-only'' theory must have $\overline{c}=2$. Hence, we conclude that the combined theory is $\mathbb{Z}_3\times\overline{\mathbb{Z}_2^{(1/2)}\times\mathbb{Z}_2^{(1/2)}}|_R$, where the restriction indicates that only excitations local with respect to the electron are allowed (more on this in Sec.~\ref{secsu2su2}).
When combined with the $\mathbb{Z}_3$-parafermion chiral CFT at the edge of the SFib state, this theory for the parent medium leads to the theory $\mathrm{Fib}\times\overline{\mathbb{Z}_2\times\mathbb{Z}_2}|_R$, or equivalently $SU(2)_3\times\overline{SU(2)}_1|_R$ for the SFib to vacuum edge.

There is one additional subtlety, however, which is that the SFib phase is superconducting.
Hence, Bogoliubov-de Gennes (BdG) quasiparticles,
rather than electrons, are the quasiparticles of a superconductor with (possibly mobile) vortices.
Thus, the ``electron'' of the previous paragraph should actually be interpreted as
a BdG quasiparticle. Our previous statements are unchanged but must be supplemented
by the fact that a $-1$ results when a BdG quasiparticle encircles an $hc/2e$ vortex.

\section{CS theory of topological superconductors with $SU(2)\times SU(2)$ gauge symmetry}\label{secsu2su2}

Combining the preceding elements, we postulate a bulk low-energy effective field theory
for the electronic SFib state that takes the form of an
$SU(2)_3\times\overline{SU(2)}_1|_R$ Chern-Simons gauge theory:
\begin{equation}
\label{eqn:Fib-electron-effective-action}
{\cal L} = {\cal L}_\text{top} + {\cal L}_\text{SC} + {\cal L}_\text{qp}
\end{equation}
In this expression,
\begin{multline}
\label{eqn:Fib-electron-top-action}
 {\cal L}_\text{top} =\frac{3}{4\pi}\epsilon^{\mu\nu\lambda} \text{Tr}\left(a_\mu \partial_\nu a_\lambda + \frac{2}{3} a_\mu a_\nu a_\lambda \right)\\
- \frac{1}{4\pi}\epsilon^{\mu\nu\lambda} \text{Tr}\left(b_\mu \partial_\nu b_\lambda + \frac{2}{3} b_\mu b_\nu b_\lambda \right),
\end{multline}
Here, $a_\mu$ and $b_\mu$ are SU(2) gauge fields. Each is
a triplet of gauge fields, corresponding to the three generators of SU(2);
we will denote this triplet with an arrow: $\vec{a}_\mu, \vec{b}_\mu$.
`Tr' denotes the trace in the fundamental representation of SU(2),
which is taken by forming the $2\times 2$ matrices $a_\mu =\frac{1}{2} \vec{\tau} \cdot \vec{a}_\mu$,
$b_\mu =\frac{1}{2} \vec{\tau}\cdot  \vec{b}_\mu$ for Pauli matrices $\vec{\tau}$.
The second term in Eq. (\ref{eqn:Fib-electron-effective-action}) describes superconductivity in this system:
\begin{equation}
\label{eqn:Fib-electron-SC-action}
 {\cal L}_\text{SC} = \frac{2}{2\pi} A_\mu \epsilon_{\mu\nu\lambda} \partial_\nu c_\nu + \frac{1}{2\rho_s}  (\partial_\mu c_\nu - \partial_\nu c_\mu)^2 ,
\end{equation}
The coefficient $2/2\pi$ denotes the fact that a single flux quantum of $c_\mu$
is a Cooper pair or, equivalently, that the minimal vortex in a superconductor has magnetic flux $hc/2e=\pi$.
The coefficient of the Maxwell term is the superfluid density $\rho_s$.
This term is subleading compared to the other terms in the action and will generally be dropped
in the following discussion.
Quasiparticles carrying the
spin-$j$ representation of SU(2) are minimally coupled to the gauge fields in
the combination $\vec{a}_\mu \cdot \vec{T}_j$, where $\vec{T}_j$ are the generators of the spin-$j$
representation of the Lie algebra of SU(2):
\begin{equation}
 {\cal L}_\text{qp} = \sum_{m,n,l} {\cal L}_{m,n,l} - V(\Phi)
\end{equation}
\begin{multline}
\label{eqn:Fib-electron-qp-action}
{\cal L}_{m,n,l} = \\
\bar{\Phi}^{(\frac{m}{2},\frac{n}{2})}_l \!
\left(i\partial_0 + \vec{a}_0 \shortdot \vec{T}_\frac{m}{2}\!\otimes\! I + \vec{b}_0 \shortdot I\!\otimes\!\vec{T}_{\frac{n}{2}}
+ l{c_0} I\!\otimes\! I \right)\! \Phi^{(\frac{m}{2},\frac{n}{2})}_l
\\
-\left|\left(i\nabla_j + \vec{a}_j \shortdot \vec{T}_\frac{m}{2}\!\otimes\! I +
\vec{b}_i \shortdot I\!\otimes\!\vec{T}_{\frac{n}{2}}+ l{c_i} I\!\otimes\! I \right)\Phi^{(\frac{m}{2},\frac{n}{2})}_l \right|^2.
\end{multline}
$I$ is the identity matrix. The matter fields $\Phi^{(\frac{m}{2},\frac{n}{2})}_r$ have two indices (suppressed), one for each representation of $SU(2)$.
Thus, they can be viewed as matrix fields (in general, rectangular). The two SU(2) Lie algebras act with generators
$\vec{T}_{\frac{m}{2}}\otimes I$ and $I\otimes\vec{T}_{\frac{n}{2}}$. For compactness, we will henceforth write
$\vec{T}_{\frac{m}{2}}\equiv\vec{T}_{\frac{m}{2}}\otimes I$
and $\vec{S}_{\frac{n}{2}}\equiv I\otimes\vec{T}_{\frac{n}{2}}$.
The vorticity $l$ can be an arbitrary integer. The allowed values of $m,n$ are determined as follows.
In the unrestricted theory, braiding a particle with quantum numbers $({j_1},{j_2},l)$ around a $(3/2,1/2,0)$
quasiparticle will result in a phase $2\pi({j_1}-{j_2})$. Hence, $(3/2,1/2,0)$ quasiparticles are fermions that have Abelian braiding with all other quasiparticles. We identify this as the BdG quasiparticle. Because the BdG quasiparticle is adiabatically connected to the three-dimensional electron, we require that all allowed excitations without a vortex have mutual statistics 1 with this excitation. Likewise, all excitations with an odd number of vortices will have mutual statistics -1 with the BdG quasiparticle.
As a consequence, all other allowed quasiparticles with quantum numbers $({j_1},{j_2},l)$ must have mutual statistics $(-1)^l$ with BdG $(3/2,1/2,0)$ quasiparticles.
Therefore, $2{j_1}+2{j_2}+l$ must be even for allowed excitations. For example, flux $hc/2e$ vortices must have quantum numbers $(0,1/2,1)$, $(3/2,0,1)$, or $(1,1/2,1)$ while
topological excitations without flux must have quantum numbers $(3/2,1/2,0)$, $(1/2,1/2,0)$
or $(1,0,0)$. This last quasiparticle type carries spin-$1$ for $a_\mu$ gauge transformations
and spin-$0$ for $b_\mu$ gauge transformations; hence they are coupled to the former via $\vec{T}_1$ and
are uncoupled from the latter.

In the remainder of the paper, we shall take a perspective dual to the one in Sec.~\ref{Sec-CC}, taking the
(proposed) Chern-Simons theory (\ref{eqn:Fib-electron-effective-action}) as our starting point
and showing that the Chern-Simons theory of the parent medium may be recovered
via a symmetry-breaking transition in which one of the quasiparticles of the theory condenses.
This condensate destroys the superconducting order parameter if $l\neq0$
and (with some exceptions) leads to an Abelian state. In particular, the symmetry-breaking transition that we find between the $SU(2)_3\times\overline{SU(2)}_1|_R$ CS theory and the $\nu=2/3$ FQH state lends credence to the idea of a superconducting Fibonacci state built by proximity induced superconductivity
in the $\nu=2/3$ state.\cite{Mong14a, Stoudenmire15}


Before considering condensation transitions from the effective theory (\ref{eqn:Fib-electron-effective-action}),
we note that it generalizes to a family of superconducting states in which the
Bogoliubov-de Gennes quasiparticles fractionalize so that an $SU(2)\times SU(2)$
local gauge symmetry emerges. These theories have a fermionic particle that braids trivially with all of the other particles; we identify it with the electron.
At the edge, these theories have an $SU(2)$ Kac-Moody elgebra associated with spin symmetry and an $SU(2)$ Kac-Moody algebra that is associated with
particlehole symmetry -- the latter is analogous to a Landau-Ginzburg theory for paired quantum Hall states \cite{Fradkin99b} and is also manifested in a coupled
wire construction of those states \cite{Teo14}.
The levels $3$ and $-1$ of the SU(2) Chern-Simons fields in Eq. (\ref{eqn:Fib-electron-effective-action}) can be generalized
to $p+q$ and $p-q$. So long as $p$ is odd, we are guaranteed to have a spin-$1/2$ fermion in the theory.
If we denote an excitation of the theory carrying the $m/2$ representation of $SU(2)_{p+q}$ and the $n/2$ representation of $SU(2)_{p-q}$ as $\Phi^{(\frac{m}{2},\frac{n}{2})}$,
then it is consistent to identify $\Phi^{\left((p+q)/2,(p-q)/2\right)}$ as the BdG quasiparticle.
(We label representations of counterpropagating $SU(2)$ components with negative indices for notational convenience.)
Locality with respect to the BdG quasiparticle leads to the restriction that $m+n$ must be an even number. Of course, we can attach flux $lhc/2e$ to such
an excitation, thereby leading to a phase of $(-1)^l$ upon encircling a BdG quasiparticle and the somewhat loosened restriction:
\begin{equation}
  m+n+l\equiv0 \mod 2.
\end{equation}
We call the theory itself $SU(2)_{p+q}\times SU(2)_{p-q}\vert_R$ to indicate the restriction.\footnote{It is an important property of $SU(2)_{p+q}\times SU(2)_{p-q}$ that the restriction $m+n+l$ even is compatible with the fusion rules of the theory.}
This more general theory takes the form:
\begin{equation}
{\cal L} = {\cal L}_\text{top} + {\cal L}_\text{SC} + {\cal L}_\text{qp}
\end{equation}
where we now have
\begin{multline}
\label{eqn:general-top-action}
 {\cal L}_\text{top} =\frac{p+q}{4\pi}\epsilon^{\mu\nu\lambda} \text{Tr}\left(a_\mu \partial_\nu a_\lambda + \frac{2}{3} a_\mu a_\nu a_\lambda \right)\\
- \frac{q-p}{4\pi}\epsilon^{\mu\nu\lambda} \text{Tr}\left(b_\mu \partial_\nu b_\lambda + \frac{2}{3} b_\mu b_\nu b_\lambda \right),
\end{multline}
\begin{multline}
\label{eqn:general-qp-action}
 {\cal L}_\text{qp} = \sum_{m,n,l}\left[ \bar{\Phi}^{(\frac{m}{2},\frac{n}{2})}_l \left(i\partial_0 + g_0^{mnl}\right)
\Phi^{(\frac{m}{2},\frac{n}{2})}_l\right.\\
-\left. \left|\left(i\nabla_j + g_j^{mnl} I\right)\Phi^{(\frac{m}{2},\frac{n}{2})}_l \right|^2 \right]
- V(\{\Phi^{(\frac{j}{2},\frac{k}{2})}_r \}),
\end{multline}
and ${\cal L}_\text{SC}$ is unchanged from Eq. (\ref{eqn:Fib-electron-SC-action}).
The potential $V(\{\Phi^{(\frac{j}{2},\frac{k}{2})}_r \})$ depends on the full set of the quasiparticle
fields $\{\Phi^{(\frac{j}{2},\frac{k}{2})}_r \}$ and can couple the different quasiparticle fields,
subject to consistency with the fusion rules.
In this equation, $g_\mu^{mnl}$ denotes
\begin{equation}
g^{mnl}_\mu=\vec{a}_\mu \cdot \vec{T}_{\frac{m}{2}} +
\vec{b}_\mu \cdot \vec{S}_{\frac{n}{2}} + l{c_\mu} I,
\end{equation}
As a connection to more familiar systems, we note that the usual $p+ip$ superconductor may be thought of as lying in this class of topological superconducting states, with $p=q=1$ in that case yielding $SU(2)_2$ anyons for which the twist field may only appear in the presence of a magnetic flux, i.e., in the core of a superconducting vortex.\cite{Read00}
In the following sections, we will consider transitions from this topological superconducting state to other topological states,
both Abelian and non-Abelian.

\section{Pedagogical Example: Transition From Doubled Ising to Toric Code}

In this section, we work out in full detail a simple pedagogical example that illustrates the basic features of the general class of transitions discussed in later sections. We consider the doubled Ising(-like) theory that is the $p=0,q=2$ instance of
the effective theories of the previous section. The coupling to the external gauge field is inessential to this discussion,
so we will set ${\rho_s}=0$ and ignore the external gauge field and the vorticity of the matter fields.

This theory has $9$ matter fields $\Phi^{(\frac{m}{2},\frac{n}{2})}$ with $m,n=0,1,2$.
We focus on the condensation of $\Phi^{(1,1)}$, which is the only condensate that can lead to
a non-trivial phase. When $\langle\Phi^{(1,1)}\rangle =\Phi \neq 0$,
${\cal L}_\text{qp}$ in Eq. (\ref{eqn:Fib-electron-qp-action}) takes the following
form at low energy:
\begin{multline}
\label{eqn:example-condensed action}
 {\cal L}_\text{qp} =
 \left|\left(\vec{a}_j \shortdot \vec{T}_1 +
\vec{b}_i \shortdot \vec{S}_{1} \right) \! \Phi\right|^2\\
+
{\sum_{m,n}}' \left[ \bar{\Phi}^{(\frac{m}{2},\frac{n}{2})}
\left(i\partial_0 + \vec{a}_0 \cdot \vec{T}_\frac{m}{2} +
\vec{b}_0 \cdot \vec{S}_{\frac{n}{2}}
\right)\Phi^{(\frac{m}{2},\frac{n}{2})}
\right.\\
-\left. \left|\left(i\nabla_j + \vec{a}_j \cdot \vec{T}_\frac{m}{2} +
\vec{b}_i \cdot \vec{S}_{\frac{n}{2}} \right)\Phi^{(\frac{m}{2},\frac{n}{2})} \right|^2\right]- V(\Phi).
\end{multline}
The prime indicates that $m=n=1$ is not included in the summation.
Here, we have taken the gauge $a_0 = b_0 = 0$, so we must also enforce the corresponding Chern-Simons
constraint equations:
\begin{eqnarray}
\label{eq-example-fluxattach}
  \sum_{mn}\bar{\Phi}^{(\frac{m}{2},\frac{n}{2})} \vec{T}_{\frac{m}{2}}\Phi^{(\frac{m}{2},\frac{n}{2})}
&+&\frac{2}{4\pi}\epsilon^{0ij}\left(\partial_i\vec{a}_j+\frac{1}{2}\vec{a}_i\times\vec{a}_j\right)=0\nonumber\\
  \sum_{mn}\bar{\Phi}^{(\frac{m}{2},\frac{n}{2})} \vec{S}_{\frac{n}{2}}\Phi^{(\frac{m}{2},\frac{n}{2})}
&-&\frac{2}{4\pi}\epsilon^{0ij}\left(\partial_i\vec{b}_j+\frac{1}{2}\vec{b}_i\times\vec{b}_j\right)=0\nonumber\\
\end{eqnarray}

We assume that $\Phi = \phi (0,1,0)\otimes(0,1,0)$ in a basis for the spin-$1$ representation
of SU(2) spanned by eigenvectors of $T^z$ with eigenvalues $1,0,-1$, so that $T^z = \text{diag}(1,0,-1)$,
where, as usual, the SU(2) generators $T^z$, $T^{\pm}$ satisfy the commutation relations
$[{T^z},T^{\pm}]=\pm T^{\pm}$ and $[{T^+},{T^-}]=2 T^z$.
In this notation, we write $\vec{a}_j \shortdot \vec{T}_1 =
a^z_j \shortdot {T}_1^z + a^+_j \shortdot {T}_1^- + a^-_j \shortdot {T}_1^+$ and similarly for $\vec{b}_j$.
The first term in Eq. (\ref{eqn:example-condensed action}) can now be written in the form:
\begin{equation}
\label{eqn:gauge-mass-example}
\phi^2 \left(a^+_j  a^-_j + b^+_j  b^-_j \right)
\end{equation}
Thus, it is a mass term for the $\pm$ SU(2) components of the gauge fields. The $z$ components are
left massless, and the remaining unbroken gauge symmetry is $U(1)\times U(1)\rtimes \mathbb{Z}_2$.
The extra discrete $\mathbb{Z}_2$ gauge symmetry is the invariance of the condensate under a $\pi$ rotation
about the $x$-axis in SU(2), which takes $a^z_j\rightarrow -a^z_j$.

Any matter field that is a source (via the constraint (\ref{eq-example-fluxattach}))
for a massive gauge field must be confined since the energy cost to separate two such quasiparticles
is linear in the separation. These are the matter fields that braid non-trivially with the condensate:
only $a^\pm$ and $b^\pm$ couple to the condensate, as in Eq. (\ref{eqn:gauge-mass-example}),
so matter fields whose density is locked to the flux of these gauge fields by the Chern-Simons constraints
braid non-trivially with the condensate. The only fields that are left deconfined are $\Phi^{(0,0)}$, $\Phi^{(1,0)}$,
and $\Phi^{(\frac{1}{2},\frac{1}{2})}$. The field $\Phi^{(0,1)}$ is equivalent to $\Phi^{(1,0)}$ since they differ by the condensate.
The field $\Phi^{(1,0)}$ is a triplet under the $\vec{a}_j$ SU(2). When the symmetry is broken down to U(1),
the triplet splits into three different fields, with $a_j$ charges $q_a = 0,\pm 1$. The charge-0 field decouples from the gauge fields
and is topologically trivial.  The field $\Phi^{(\frac{1}{2},\frac{1}{2})}$ is a doublet under both SU(2) gauge symmetries.
When they break down to $U(1)\times U(1)$, the (doublet)$^2$ splits into 4 fields, with charges ${q_a},{q_b}=\pm 1/2$
for. respectively, $a_j$ and $b_j$.

By gapping some of the gauge fields, we alter the braiding statistics of the remaining matter fields.
The effective action takes the form:
\begin{multline}
\label{eqn:example-abelian-action}
 {\cal L}_\text{qp} =
{\sum_{{q_a},{q_b}}} \left[ \bar{\Phi}^{({q_a},{q_b})}
\left(i\partial_0 + {q_a} a_0 +
{q_b} b_0
\right)\Phi^{({q_a},{q_b})}
\right.\\
-\left. \left|\left(i\nabla_j + {q_a} a_j +
{q_b} b_i  \right)\Phi^{({q_a},{q_b})} \right|^2\right] - V(\Phi)\\
+ \frac{1}{4\pi}\epsilon^{i0j} a_i \partial_0  a_j
- \frac{1}{4\pi}\epsilon^{i0j} b_i \partial_0  b_j
\end{multline}
The charges take values $({q_a},{q_b}) = (\pm 1,0), (\pm 1/2, \pm 1/2),  (\pm 1/2, \mp 1/2)$.
We have dropped the $z$ superscripts from the gauge fields: $a_i\equiv a_i^z$, $b_i\equiv a_i^z$.
Note that an extra factor of $1/2$ has appeared in front of the Chern-Simons terms due to the normalization
of the Trace defined after Eq. (\ref{eqn:Fib-electron-top-action}). It is more natural to normalize U(1) Chern-Simons gauge fields
so that charges take integral values, rather than the half-integral values that $T^z$ eigenvalues take. Thus, we define
$a^{1,2} \equiv a \pm b$, in terms of which the action takes the form:
 \begin{multline}
\label{eqn:example-abelian-action-simplified}
 {\cal L}_\text{qp} =
{\sum_{m,n}} \left[ \bar{\Phi}^{({q_1},{q_2})}
\left(i\partial_0 + {q_1} a^1_0  +
{q_1} a^2_0
\right)\Phi^{({q_1},{q_2})}
\right.\\
-\left. \left|\left(i\nabla_j + {q_1} a^1_j  +
{q_1} a^2_i \right)\Phi^{({q_1},{q_2})} \right|^2\right] - V(\Phi)\\
+ \frac{2}{2\pi}\epsilon^{i0j} a^1_i \partial_0  a^2_j
\end{multline}
The charges take values $({q_1},{q_2}) = (1,0), (0,1), (1,1)$.
This is the effective theory for the Toric Code or deconfined phase of $\mathbb{Z}_2$ gauge theory, as expected from the analysis of Bais and Slingerland \cite{Bais09}. However, this appears to be simply a $U(1)\times U(1)$ gauge theory. What happened to the $\rtimes\mathbb{Z}_2$ portion of the theory? The $\mathbb{Z}_2$ action of the theory would interchange $a^1$ and $a^2$. In principle, gauging this action would lead back to and $\mathrm{Ising}\times\mathrm{Ising}$ theory.\cite{Barkeshli14} In this case, however, the generator of the $\mathbb{Z}_2$ part of the theory must create a twist in the $a$ field without creating one in the $b$ field (or vice versa). It is therefore a confined excitation, coming from the field $\Phi^{(\frac{1}{2},0}$ (or $\Phi^{0,\frac{1}{2}}$) of the original theory. The theory of Eq.~(\ref{eqn:example-abelian-action}) may thus be taken at face value, and the resulting phase is topologically equivalent to the Toric Code. In the next section, we generalize the preceding discussion.

\section{Condensation Transitions from $SU(2)_{p+q}\otimes SU(2)_{p-q}\vert_R $}
\label{Sec:e-condense}

We consider the condensation of a field $\Phi^{(\frac{m}{2},\frac{n}{2})}_l$, which carries the $m/2$ representation of $SU(2)_{p+q}$ and the $n/2$ representation of $SU(2)_{p-q}$, along with $l$ units of attached magnetic flux. Different choices of $(m,n,l)$ cause transitions
to different states. While some choices will lead to trivial insulating states, other choices will lead to interesting states with a different
type of topological order and/or superconductivity. These different possibilities can be analyzed as follows.

We may make the gauge choice $\vec{a}_0=0$, $\vec{b}_0=0$ so long as we also include the restrictions coming from the $\vec{a}_0$ and $\vec{b}_0$ equations of motion. Because $\vec{a}_0$ and $\vec{b}_0$ only appear linearly in the Lagrangian, they act as Lagrange multipliers for the attachment of non-Abelian gauge flux to the quasiparticle $\Phi^{(\frac{m}{2},\frac{n}{2})}_l$. That is,
\begin{eqnarray}\label{eq-fluxattach}
  \sum_{mnl}\bar{\Phi}^{(\frac{m}{2},\frac{n}{2})}_l \vec{T}_{\frac{m}{2}}\Phi^{(\frac{m}{2},\frac{n}{2})}_l&+&\frac{p+q}{4\pi}\epsilon^{0ij}\left(\partial_i\vec{a}_j+\frac{1}{2}\vec{a}_i\times\vec{a}_j\right)=0\nonumber\\
  \sum_{mnl}\bar{\Phi}^{(\frac{m}{2},\frac{n}{2})}_l \vec{S}_{\frac{n}{2}}\Phi^{(\frac{m}{2},\frac{n}{2})}_l&+&\frac{p-q}{4\pi}\epsilon^{0ij}\left(\partial_i\vec{b}_j+\frac{1}{2}\vec{b}_i\times\vec{b}_j\right)=0\nonumber\\
\end{eqnarray}
Likewise, we can make the gauge choice $c_0=0$ accompanied by the restriction
\begin{equation}\label{flux}
  \sum_{mnl} l\bar{\Phi}^{(\frac{m}{2},\frac{n}{2})}_l\Phi^{(\frac{m}{2},\frac{n}{2})}_l+\frac{1}{\pi}\epsilon^{0ij}\partial_i A_j=0.
\end{equation}
Note that it is this last condition that guarantees that $\Phi^{(\frac{m}{2},\frac{n}{2})}_l$ carries $l$ flux quanta of the external magnetic field $A_\mu$. Importantly, if there are no quasiparticles present in the system, there is also no magnetic flux. This is the Meissner effect.

After we make these gauge choices, we are ready to consider the effect of quasiparticle condensation: we assume that the potential $V(\{\Phi^{(\frac{j}{2},\frac{k}{2})}_r \})$ is such that $\Phi^{(\frac{m}{2},\frac{n}{2})}_l$ (and no other quasiparticle field)
acquires a constant expectation value, which we will denote by $\Phi$ (we have suppressed the SU(2) indices of $\Phi$),
so that the total Lagrangian becomes
\begin{eqnarray}
{\cal L} &=&-|(\vec{a}_i\shortdot \vec{T}_{\frac{m}{2}} +\vec{b}_i\shortdot\vec{S}_{\frac{n}{2}} + l I c_i)\Phi|^2 + \frac{1}{\pi} \epsilon^{\mu\nu i} A_\mu\partial_\nu c_i\nonumber\\
&+&\frac{p+q}{8\pi}\epsilon^{i0j}\vec{a}_i\shortdot\partial_0  \vec{a}_j
+\frac{p-q}{8\pi}\epsilon^{i0j}\vec{b}_i\shortdot\partial_0  \vec{b}_j
\end{eqnarray}
The first term gives a gap to a combination of the gauge fields, by the Anderson-Higgs mechanism. The topological order of
the resulting phase is determined by the remaining gauge symmetries that leave $\Phi$ invariant and their Chern-Simons terms.
In the sections that follow, we carry out this analysis for different $m, n, l$.

\subsection{Transitions to Abelian quantum Hall States}

We first assume that $l\neq0$, $m\neq0$, and $n\neq 0$. The first condition guarantees that superconductivity is destroyed.
The survival of a subset of the gauge symmetries of the system (together with their Chern-Simons terms) leads to
the quantum Hall effect. In this subsection, we will focus on the cases in which the resulting quantum Hall state is Abelian.

To lighten the notation, we define $\bar{\Phi}\mathcal{O}\Phi\equiv\abs{\Phi}^2\expt{\mathcal{O}}$ for general operator $\mathcal{O}$. Since $l\neq0$, we may integrate out the gapped field $c_i+\frac{1}{l} \vec{a}_i\shortdot \expt{\vec{T}_{\frac{m}{2}}}+\frac{1}{l} \vec{b}_i\shortdot \expt{\vec{S}_{\frac{n}{2}}}$ to obtain
\begin{eqnarray}
{\cal L}_\text{eff} &=&-|\Phi^{(\frac{m}{2},\frac{n}{2})}_l|^2
\left[\vec{a}_i\shortdot M_{m}\shortdot \vec{a}_i+ \vec{b}_i\shortdot M_{n}\shortdot \vec{b}_i\right]\nonumber\\
 &-& \frac{1}{\pi l} \epsilon^{\mu\nu i} A_\mu \partial_\nu \left(\vec{a}_i\shortdot
\expt{\vec{T}_{\frac{m}{2}}}+\vec{b}_i\shortdot \expt{\vec{S}_{\frac{n}{2}}}\right)\nonumber\\
&+&\frac{p+q}{8\pi}\epsilon^{i0j}\vec{a}_i\shortdot\partial_0  \vec{a}_j
+\frac{p-q}{8\pi}\epsilon^{i0j}\vec{b}_i\shortdot\partial_0  \vec{b}_j.
\end{eqnarray}
The effective mass matrix \mbox{$M_m=\expt{\vec{T}_{\frac{m}{2}}\vec{T}_{\frac{m}{2}}}-\expt{\vec{T}_{\frac{m}{2}}}\expt{\vec{T}_{\frac{m}{2}}}$} (and likewise for $M_n$) depends on the representations $j=m/2,n/2$
of \mbox{$SU(2)_{p+q}\otimes SU(2)_{p-q}\vert_R$} carried by $\Phi^{(\frac{m}{2},\frac{n}{2})}_l$.
However, we may deduce some properties without referring to a specific representation.
Taking the expectation value of $M_m$ in the direction of a unit vector $\hat{v}$, we find
\begin{eqnarray}\label{eq-CS}
\hat{v}\shortdot M_m\shortdot \hat{v}&=&\expt{(\vec{T}_{\frac{m}{2}}\shortdot \hat{v})^2}-\expt{\vec{T}_{\frac{m}{2}}\shortdot\hat{v}}\expt{\vec{T}_{\frac{m}{2}}\shortdot\hat{v}}\nonumber\\
                           &=&\bracket{u}{u}\bra{u}(\vec{T}_{\frac{m}{2}}\shortdot \hat{v})^2\ket{u}-(\bra{u}\vec{T}_{\frac{m}{2}}\shortdot \hat{v}\ket{u})^2\nonumber\\
                           &\geq&0,
\end{eqnarray}
where $\ket{u}$ is the normalized internal state vector\footnote{$\ket{u}=\Phi^{(\frac{m}{2},\frac{n}{2})}_l/\abs{\Phi^{(\frac{m}{2},\frac{n}{2})}_l}$} of $\Phi^{(\frac{m}{2},\frac{n}{2})}_l$ and the last line follows from the Cauchy-Schwartz inequality. This shows that the effective mass matrix is positive semi-definite, gapping out parts of the gauge field. There is a remaining continuous gauge symmetry if and only if equality holds in the above equation for some unit vector $\hat{v}$.\footnote{Even if all eigenvalues of the mass matrix are non-zero, there may be discrete gauge symmetries that remain.} By the Cauchy-Schwartz inequality, this only occurs when
$\vec{T}_{\frac{m}{2}}\shortdot \hat{v} \ket{u} \propto \ket{u}$. For any such vector, $\hat{v}\cdot \vec{a}$ is gapless.
Similarly, for any vector $\hat{v}'$ such that $\hat{v}'\shortdot M_m\shortdot \hat{v}' = 0$, $\hat{v}'\cdot \vec{b}$ is gapless.

Focusing now on condensates that preserve a non-zero continuous gauge symmetry\footnote{Favored, perhaps, by phase space considerations.}, we set $\hat{v}$ to be in the $\hat{z}$ direction, so $\hat{v}\shortdot \vec{T}_{\frac{m}{2}}=T^z_{\frac{m}{2}}$. Further, we let $T^z_{\frac{m}{2}}\ket{u}=r_a\ket{u}$ and $S^z_{\frac{n}{2}}\ket{u}=r_b\ket{u}$ where $\abs{r_a}\leq \abs{m}/2$, $\abs{r_b}\leq \abs{n}/2$ and $r_a$ and $r_b$ are integers or half-integers depending on the representation. We assume that both $r_a$ and $r_b$ are non-zero
for the remainder of this subsection (and consider the case of vanishing $r_a$, $r_b$ in the next section).
Then, after integrating out the gapped degrees of freedom we have
\begin{eqnarray}
{\cal L}_\text{eff} &=& - \frac{1}{\pi l} \epsilon^{\mu\nu i} A_\mu \partial_\nu \left(a^z_i r_a+b^z_i r_b)\right)\nonumber\\
&+&\frac{p+q}{8\pi}\epsilon^{i0j}a^z_i\partial_0  a^z_j
+\frac{p-q}{8\pi}\epsilon^{i0j}b^z_i\partial_0  b^z_j
\end{eqnarray}
along with the constraints
\begin{eqnarray}
  r_a|\Phi^{(\frac{m}{2},\frac{n}{2})}_l|^2+\frac{p+q}{4\pi}\epsilon^{0ij}\partial_i a^z_j&=&0,\nonumber\\
  r_b|\Phi^{(\frac{m}{2},\frac{n}{2})}_l|^2+\frac{p-q}{4\pi}\epsilon^{0ij}\partial_i b^z_j&=&0,
\end{eqnarray}
and
\begin{equation}
   l|\Phi^{(\frac{m}{2},\frac{n}{2})}_l|^2+\frac{1}{\pi}\epsilon^{0ij}\partial_i A_j=0
\end{equation}
We can now eliminate the order parameter by combining the constraints and
incorporate these restrictions into the Lagrangian through the use of new Lagrange multiplier fields $a_0$ and $b_0$, defining $a_i=a^z_i$, $b_i=b^z_i$. We find the following effective action for the condensed phase:
\begin{eqnarray}
{\cal L}_\text{eff} &=& - \frac{1}{\pi l} \epsilon^{\mu\nu \lambda} A_\mu \partial_\nu \left(a_\lambda r_a+b_\lambda r_b\right)\nonumber\\
&+&\frac{p+q}{8\pi}\epsilon^{\mu\nu\lambda}a_\mu\partial_\nu  a_\lambda
+\frac{p-q}{8\pi}\epsilon^{\mu\nu\lambda}b_\mu\partial_\nu  b_\lambda.
\end{eqnarray}
Writing $a=(a^{(1)}+a^{(2)})/2$, $b=(a^{(1)}-a^{(2)})/2$, we acquire the more familiar form
\begin{equation}
\label{Leff}
{\cal L}_\text{eff}=\frac{1}{4\pi}\epsilon^{\mu\nu\lambda}a^I_\mu K_{IJ} \partial_\nu a^J_\lambda - \frac{1}{2\pi} A_\mu \epsilon^{\mu\nu\lambda} t_I\partial_\nu a^I_\lambda
\end{equation}
where
\begin{equation}
\label{eqn:K-matrix-secB}
  K=\left(\begin{array}{cc}p  & q \\ q& p\end{array}\right)
\end{equation}
and $t=2/l\left( r_a+r_b, r_a-r_b\right)$ is an integer vector (for $l=1$ or $l=2$) due to the locality restriction on $\Phi^{(\frac{m}{2},\frac{n}{2})}_l$.
The filling fraction of the resulting state is (for $l\neq0$):
\begin{equation}
\nu=\frac{8r_a^2}{l^2(p+q)}+\frac{8r_b^2}{l^2(p-q)}
\end{equation}

\subsection{Examples}
\subsubsection{$SU(2)_{3}\otimes SU(2)_{-1}\vert_R $: Fermionic Fibonacci Superconductor}

We now return our attention to the Fibonacci Superconductor described by the gauge group $SU(2)_{3}\otimes SU(2)_{-1}\vert_R $ and Lagrangian~(\ref{eqn:Fib-electron-effective-action}). Condensing the field $\Phi^{(\frac{1}{2},0)}_1$ with spin projections $r_a=1/2$, $r_b=0$ gives
\begin{equation}
  K=\left(\begin{array}{cc}1  & 2 \\ 2& 1\end{array}\right)
\end{equation}
and $t=(1,1)$ so $\nu=2/3$. To construct the fields of the resulting theory, we look to the families of simple currents of the original theory: $\Phi^{(0,0)}_{2z}$, $\Phi^{(\frac{3}{2},0)}_{2z+1}$, $\Phi^{(0,\frac{1}{2})}_{2z+1}$, and $\Phi^{(\frac{3}{2},\frac{1}{2})}_{2z}$. Of these, the second and last do not commute with the condensate. The first and the third do, however, producing charges $(-z,-z)$ and $(\frac{1}{2}-s-\frac{2z+1}{2},s-\frac{1}{2}-\frac{2z+1}{2})$ where $s\in\{0,1\}$ and $z\in\mathbb{Z}$. These two families of fields have spins $z^2/3$ and $\frac{(2z+1)^2}{12}-\frac{(1-2s)^2}{4}$. Together these give the complete set of spins (equivalent $\mathrm{mod} 1$ to $-\frac{z^2}{6}$ for $z\in\mathbb{Z}$) for the $(112)$ theory with $K$ matrix given above. This completes the inversion of the construction given in Ref.~\onlinecite{Mong14a}. We have shown a condensation transition from the fermionic Fibonacci state constructed in that work directly back to the parent $\nu=2/3$ FQH state from which it was constructed.

\subsubsection{$SU(2)_{4}\otimes SU(2)_{2}\vert_R $}
The fields of the new theory do not always descend solely from the simple currents of the old. Specifically, if the condensate carries a simple current representation of one of the $SU(2)$s, then other fields will remain deconfined in the resulting theory. For instance, if we consider the condensation of the field $\Phi^{(2,\frac{1}{2})}_1$ in $SU(2)_{4}\otimes SU(2)_{2}\vert_R $ theory, we find that two families of fields commute with the condensate: $\Phi^{(\mathrm{even}/2,0)}_{2z}$ and $\Phi^{(\mathrm{odd}/2,1)}_{2z+1}$. The first family includes all the deconfined simple currents of the theory, but only allows spins of the form $\frac{s^2}{4}+\frac{z^2}{2}$ for $s$ and $z$ integer. The second family completes the set of spins in the resultant $(331)$ theory, with spins $\frac{(2s+1)^2}{16}+\frac{(2z+1)^2}{8}$.
\subsubsection{$l>1$}
It should be noted that in cases with $l>1$ not all of the fields of the new theory descend directly from the deconfined fields of the old. Rather, some of the new fields appear as vortices in the condensate. Due to the condensation of multiple fluxes, these vortices will carry fractional charge. This is entirely in line with, e.g. the composite fermion construction of the fractional quantum Hall states\cite{Jain07} and we shall not focus further attention on it here.
\subsection{Transitions to metaplectic quantum Hall states}
We now consider the condensate of the previous subsection, but with either $r_a=0$ or $r_b=0$.
 Then the effective Lagrangian described by Eq.~(\ref{Leff}) has an additional symmetry.
For example, if $r_a=0$ then the Lagrangian is invariant under
\begin{equation}
a^I\rightarrow R^a_{IJ}a^J
\end{equation}
with
\begin{equation}
R^a=-\sigma_x=\left(\begin{array}{cc}0 & -1\\ -1& 0\end{array}\right).
\end{equation}
Because $R^a\in SU(2)\times SU(2)$, this is actually a remaining discrete gauge symmetry of the Lagrangian, enhancing the symmetry in the special case $r_a=0$ from $U(1)\otimes U(1)$ to $U(1)\otimes U(1)\rtimes \mathbb{Z}_2$. Likewise, when $r_b=0$, the symmetry is again enhanced to $U(1)\otimes U(1)\rtimes \mathbb{Z}_2$, this time by $R^b$, with
\begin{equation}
R^b=\sigma_x=\left(\begin{array}{cc}0 & 1\\ 1& 0\end{array}\right).
\end{equation}

One interesting example of such a state is given when
\begin{equation}
K=\left(\begin{array}{cc}1 & 2\\ 2& 1\end{array}\right)
\end{equation}
In this case the Lagrangian and symmetry group considered here map exactly to one of those shown by Barkeshli and Wen \cite{Barkeshli10} to support $\mathbb{Z}_4$ parafermion-type excitations.

These are examples of metaplectic states. This family of topological phases, indexed by
prime number $P>2$ has particle types $1$, $Z$; $X$, $X'$; and ${Y_1}, \ldots, Y_{s}$, where
$P=2s+1$. The special case $P=3$ is equivalent
to $SU(2)_4$, and the $X,{Y_1},X',Z$ particles correspond to
spins $\frac{1}{2},1,\frac{3}{2},2$.
The topological properties of the metaplectic TQFT are as follows \cite{Rowell12,Naidu11,Hastings13}.
The topological spins $\theta_a = e^{2\pi h_a}$ of these particles are given by
${h_I} = 0, {h_Z}=1, {h_X}=\frac{s}{8}, h_{X'}=\frac{s+4}{8}, h_{Y_j}=\frac{j(P-j)}{2P}$.
Their fusion rules are:
\begin{eqnarray}
\label{eqn:fusion}
X \cdot X &=& I + {\sum_i} {Y_i}\, , {\hskip 0.5 cm}
X \cdot X' =  Z + {\sum_i} {Y_i}\, , \cr
X \cdot Z &=& X'  \, , {\hskip 1.55 cm}
Z \cdot {Y_i} = Y_i  \, ,\cr
X \cdot {Y_i} &=& X + X' \, , {\hskip 0.8 cm}
Z \cdot Z = I \, ,\cr
Y_i \cdot Y_j &=& Y_{|i-j|} + Y_{\text{min}(i+j,P-i-j)}\, ,\mbox{for $i\neq j$}\cr
Y_i \cdot Y_i &=& I + Z + Y_{\text{min}(2i,P-2i)}
\end{eqnarray}
For the $P=3$ case, there is a single ${Y_i}$, which we will simply call $Y\equiv Y_1$,
and the last of these fusion rules is modified to $Y\cdot Y = I + Y + Z$ or,
in the notation of $SU(2)_4$, $1\times 1 = 0 + 1+ 2$.
Barkeshli and Wen \cite{Barkeshli10} showed that $\mathbb{Z}_2$ vortices in
the $U(1)\otimes U(1)\rtimes \mathbb{Z}_2$ theory correspond to $X$ particles while $Y$ particles
carry $U(1)$ flux that is anti-symmetric in the two $U(1)$ factors.
This anyon model is closely related to models of parafermionic zero modes at
defects in gapped fractional quantum Hall states \cite{Clarke13a, Lindner12, Cheng12}.
It has recently been shown that the $P=3$ case of metaplectic anyons is universal
for quantum computation when braiding is supplemented by measurement \cite{Cui15,Levaillant15};
this is likely to be true for $P>3$ as well.

\subsection{Transitions to Superconducting States}

If $l=0$, then no flux is attached to the condensed quasiparticle. In this case, the resulting state is superconducting, as can be seen from the fact that the constraint (\ref{flux}) is a statement of the Meissner effect for the electromagnetic field. The decoupling of the $U(1)$ gauge field $c_\mu$ from the $SU(2)$ fields results in an alteration of the mass matrix for those fields. Now the mass term becomes
\begin{multline}
-{\cal L}_M = |(\vec{a}_i\shortdot \vec{T}_{\frac{m}{2}} +\vec{b}_i\shortdot\vec{S}_{\frac{n}{2}})\Phi|^2 \\
    = |\Phi|^2\!\!\left[\vec{a}_i\shortdot M_m\shortdot\vec{a}_i +\vec{b}_i\shortdot M_n\shortdot\vec{b}_i
\right.\\ \left. +\left(\vec{a}_i\shortdot\!\expt{\vec{T}_{\frac{m}{2}}}+\vec{b}_i
\shortdot\!\expt{\vec{S}_{\frac{n}{2}}}\right)^2\right],
\end{multline}
where \mbox{$M_m=\expt{\vec{T}_{\frac{m}{2}}\vec{T}_{\frac{m}{2}}}-
\expt{\vec{T}_{\frac{m}{2}}}\expt{\vec{T}_{\frac{m}{2}}}$} as before.

For any remaining massless mode, all three of the terms above must vanish, since each is individually positive semi-definite. This allows us to use the usual Cauchy-Schwartz argument surrounding Eq.~(\ref{eq-CS}) to set
\begin{eqnarray}
T^z_{\frac{m}{2}}\ket{u}&=&r_a\ket{u},\nonumber\\
S^z_{\frac{n}{2}}\ket{u}&=&r_b\ket{u}
\end{eqnarray}
without loss of generality. This results in a mass term
\begin{equation}
-L_M=\left(r_a\vec{a}_i^z+r_b\vec{b}_i^z\right)^2\geq 0.
\end{equation}
If $r_a$ or $r_b$ are non-zero, this mass term results in the reduction of the symmetry of the theory from $U(1)\times U(1)$ to $U(1)$. Our Lagrangian is
\begin{eqnarray}
{\cal L}_\text{eff} &=& - \frac{1}{\pi} \epsilon^{\mu\nu \lambda} A_\mu \partial_\nu c_\lambda-\left(r_a\vec{a}_i^z+r_b\vec{b}_i^z\right)^2\nonumber\\
&+&\frac{p+q}{8\pi}\epsilon^{i0j}a^z_i\partial_0  a^z_j+\frac{p-q}{8\pi}\epsilon^{i0j}b^z_i\partial_0  b^z_j
\end{eqnarray}
along with the constraints
\begin{eqnarray}
  ir_a|\Phi|^2+\frac{p+q}{4\pi}\epsilon^{0ij}\partial_i a^z_j&=&0\nonumber\\
  ir_b|\Phi|^2+\frac{p-q}{4\pi}\epsilon^{0ij}\partial_i b^z_j&=&0,
\end{eqnarray}
which may be rearranged to find
\begin{equation}
  i(r_a^2(p-q)+r_b^2(p+q))|\Phi|^2+\frac{p^2-q^2}{4\pi}\epsilon^{0ij}\partial_i (r_a a^z_j+r_b b^z_j)=0
\end{equation}
Since the field $r_a a^z_j+r_b b^z_j$ is gapped, $\frac{p^2-q^2}{4\pi}\epsilon^{0ij}\partial_i (r_a a^z_j+r_b b^z_j)$ cannot take on a constant finite value without costing an infinite amount of energy. Therefore, since $|\Phi|^2>0$, we must have that either $p^2=q^2$ or
\begin{equation}\label{eq-rrestrict}
  r_a^2(p-q)+r_b^2(p+q)=0.
\end{equation}
We see that it is impossible to have a condensed superconducting state of this type if $\abs{p}\geq \abs{q}$ unless $r_a=r_b=0$.\footnote{If $p^2=q^2$, then one of the $SU(2)$s is trivial and the same arguments lead to $r=0$ in the other.} If $\abs{p}<\abs{q}$ we have other possible solutions (e.g. $p=3$, $q=5$, $r_a=\pm 2r_b$). In these cases the effective Lagrangian takes the form
\begin{eqnarray}
{\cal L}_\text{eff} &=& - \frac{1}{\pi} \epsilon^{\mu\nu \lambda} A_\mu \partial_\nu c_\lambda\nonumber\\
&+&\left(\frac{p+q}{8\pi}+\frac{p-q}{8\pi}\frac{r_a^2}{r_b^2}\right)\epsilon^{i0j}a^z_i\partial_0  a^z_j
\end{eqnarray}
after integrating out the massive field $r_a a^z_j+r_b b^z_j$. Using the constraint (\ref{eq-rrestrict}) we see that the Lagrangian actually becomes completely trivial in these cases, leaving only the Meissner term $- \frac{1}{\pi} \epsilon^{\mu\nu \lambda} A_\mu \partial_\nu c_\lambda$.
Thus, the system is an ordinary $s$-wave superconductor.
The $a^z_\mu$ gauge field has no Chern-Simons term and is confining (since it is a compact U(1) gauge field).

The more interesting case occurs when $r_a=r_b=0$. Now the system retains its $U(1)\otimes U(1)$ symmetry with Lagrangian
\begin{eqnarray}
{\cal L}_\text{eff} &=& - \frac{1}{\pi} \epsilon^{\mu\nu \lambda} A_\mu \partial_\nu c_\lambda\nonumber\\
&+&\frac{p+q}{8\pi}\epsilon^{i0j}a^z_i\partial_0  a^z_j+\frac{p-q}{8\pi}\epsilon^{i0j}b^z_i\partial_0  b^z_j
\end{eqnarray}
and constraints
\begin{eqnarray}
  ir_a|\Phi|^2+\frac{p+q}{4\pi}\epsilon^{0ij}\partial_i a^z_j&=&0\nonumber\\
  ir_b|\Phi|^2+\frac{p-q}{4\pi}\epsilon^{0ij}\partial_i b^z_j&=&0.
\end{eqnarray}
In addition, there is a leftover gauge symmetry associated with the transformations $a^z_i\rightarrow-a^z_i$ and $b^z_i\rightarrow-b^z_i$. These corresponded to $\pi$ rotations around the $x$ or $y$ axes in the original $SU(2)$ symmetries. The flux associated with this gauge symmetry is likewise inherited from the $SU(2)$ symmetries. The operator that inserts a $\pi$ gauge flux for the $a$ field is a quasiparticle operator $\Phi^{(\frac{m}{2},\frac{n}{2})}_l$ with $m$ odd. Likewise an operator with $n$ odd inserts a $\pi$ flux for the $b$ field.  One can see this relation from Eq.~(\ref{eq-fluxattach}). The flux insertion operators for the gauge transformations that take $a^z_i\rightarrow-a^z_i$ and $b^z_i\rightarrow-b^z_i$ are therefore descended from quasiparticle operators with $m$ and $n$ odd, respectively.

We know that since $r_a=r_b=0$, both $m$ and $n$ for our \emph{condensing} quasiparticle must be even.\footnote{That they have the same parity is consistent with $l=0$.} Therefore, the simple currents $\Phi^{(\frac{p+q}{2},0)}_{21+1}$, $\Phi^{(0,\frac{p-q}{2})}_{2l+1}$, and
$\Phi^{(\frac{p+q}{2},\frac{p-q}{2})}_{2l}$ will remain unconfined. If $p+q$ is odd, then former two act as the twist fields in a  metaplectic anyon theory while the latter is a composite of the two types of metaplectic anyons.  Note that only this latter one is available in the absence of flux, while the first two are necessarily bound to superconducting vortices. On the other hand, if $p+q$ is even, the only case in which any of the potential twist fields are left deconfined is when we are condensing the simple current $\Phi^{(\frac{p+q}{2},\frac{p-q}{2})}_{2l}$. In this case, fields $\Phi^{(\frac{\mathrm{odd}}{2},\frac{\mathrm{odd}}{2})}_{2l}$ act as the deconfined composite twist, while individual twists are confined. If we condense any other field than $\Phi^{(\frac{p+q}{2},\frac{p-q}{2})}_{2l}$ in a theory with $p+q$ even, we are left with a superconducting $U(1)\times U(1)$ state with no metaplectic anyons.

\subsection{Non-Abelian $SU(2)\times U(1)$ states}

In our analysis thus far, we have assumed that the condensing quasiparticle carries a non-trivial representation of both $SU(2)$ gauge groups.
However, if one of the representations is trivial, the resulting symmetry will be $U(1)\otimes SU(2)$ (or $U(1)\otimes SU(2)\rtimes \mathbb{Z}_2$ in the metaplectic case). Suppose, for instance, that the field
$\Phi^{(\frac{m}{2},0)}_l$ condenses. Then, the remaining theory is a $U(1)\otimes SU(2)_{p-q}$ theory
(possibly semidirect product with $\mathbb{Z}_2$, depending on the condensate).
However, if $p-q=1$, then the SU(2) sector of the theory is actually Abelian, and the total theory can still be expressed in
terms of a $K$-matrix, which would be the same as in Eq. (\ref{eqn:K-matrix-secB}).

\section{General Group}

\subsection{Basic Formalism}

The condensation process described in Sec.~\ref{Sec:e-condense} may be generalized to Chern-Simons theories based on arbitrary Lie groups.
The full (non-relativistic) Lagrangian for such a system is assumed to be
\begin{eqnarray}
\label{eqn:CS-LG-general-group}
L &=&\sum_{ml}\left[ i\bar{\Phi}^{m}_l\left(\partial_0 + \vec{a}_0\shortdot \rho_m(\vec{g})+I l c_0\right)\Phi^{m}_l\right.\nonumber\\
&-&\left.|\left(\nabla_i + \vec{a}_i\shortdot \rho_m(\vec{g})+I l c_i\right)\Phi^m_l|^2
- V(\abs{\Phi^{m}_l})\right]\nonumber\\
&+& \frac{1}{\pi} \epsilon^{\mu\nu\lambda}A_\mu \partial_\nu c_\lambda +k L_{CS}(\vec{a}),
\end{eqnarray}
where $\vec{g}$ are the generators of the Lie algebra $\mathcal{G}$ and $\rho_m$ is a representation of the Lie algebra. The first term in the third line of Eq. (\ref{eqn:CS-LG-general-group}) assumes that we have a superconducting condensate of charge $2$ bosons.
In a system in which the basic microscopic constitutents are bosons, we must keep this in mind
when, for instance, computing boson conductivities.
The general Chern-Simons Lagrangian is given by
\begin{equation}
\label{eqn:CS-general-group}
  L_{CS}(\vec{g})=\frac{1}{4\pi}\epsilon^{\mu\nu\lambda}\,\mathrm{tr}\left[a_\mu \partial_\nu  a_\lambda +\frac{2}{3} a_\mu a_\nu a_\lambda \right].
\end{equation}
Here, $a_\mu \equiv \vec{a}_\mu \cdot \rho_\text{fund}(\vec{g})$ so that
the trace is taken in the adjoint representation.

It is simplest to work in the gauge where \mbox{$\vec{a}_0=0$}. The associated Chern-Simons constraint attaches non-Abelian gauge flux to the quasiparticle $\Phi^{m}_l$ as
\begin{equation}\label{eq-grestrict}
  \sum_{ml}i\bar{\Phi}^{m}_l\rho_m(\vec{g})\Phi^{m}_l+\frac{k}{2\pi}\epsilon^{0ij} \mathrm{tr}\left[\vec{g}(\partial_i\vec{a}_j\shortdot\vec{g})+\vec{g}(\vec{a}_i\shortdot \vec{g}) (\vec{a}_j\shortdot\vec{g})\right]=0
\end{equation}
Similarly, the  Chern-Simons constraint  arising from the $c_0$ equation of motion is
\begin{equation}\label{eq-gflux}
  \sum_{mnl}i l\bar{\Phi}^m_l\Phi^m_l+\frac{1}{\pi}\epsilon^{0ij}\partial_i A_j=0.
\end{equation}
So long as this constaint is enforced, we may set $c_0=0$.

Following the process in Sec.~\ref{Sec:e-condense}, we assume that a single quasiparticle type \mbox{$\Phi=\Phi^{m}_l$} acquires a constant non-zero expectation value. Ignoring gapped fluctuations in $\Phi$, we acquire the symmetry-breaking Lagrangian
\begin{eqnarray}
L &=&-|(\vec{a}_i\shortdot \rho(\vec{g})+ l I c_i)\Phi|^2 - V(\Phi)
\nonumber\\&+&\frac{k I_{ad}}{4\pi}\epsilon^{i0j}\vec{a}_i\shortdot B \cdot \partial_0  \vec{a}_j+\frac{1}{\pi} \epsilon^{\mu\nu i}A_\mu \partial_\nu c_i
\end{eqnarray}
Once again, we assume that $l\neq0$, $m\neq0$.\footnote{Similar special cases to those described in Sec.~\ref{Sec:e-condense} will apply here as well. $l=0$ results in a state that remains superconducting after the symmetry breaking, while $m=0$ results in an additional discrete gauge symmetry.} The $U(1)$ field $c_i+\frac{1}{l} \vec{a}_i\shortdot \expt{\rho(\vec{g})}$ is clearly gapped. Integrating out this field leads to an effective mass matrix
\begin{equation}
 M_{\rho(\vec{g})}=\expt{\rho(\vec{g})\rho(\vec{g})}-\expt{\rho(\vec{g})}\expt{\rho(\vec{g})}
\end{equation}
for the remaining gauge fluctuations. This mass matrix is dependent upon the representation of $\mathcal{G}$ carried by $\Phi^m_l$. Again, however, we may use the Cauchy-Schwartz argument of Eq.~(\ref{eq-CS}) to show that the effective mass matrix is positive semi-definite, gapping out parts of the gauge field. There is a remaining gauge symmetry if and only if $\hat{v}\shortdot \rho(\vec{g})\ket{u}\propto\ket{u}$, where $\ket{u}$ is the normalized internal state vector\footnote{$\ket{u}=\Phi^{m}_l/\abs{\Phi^{m}_l}$} of $\Phi^{m}_l$. That is, there is a remaining gauge symmetry if and only if $\ket{u}$ is an eigenstate of $\hat{v}\shortdot \rho(\vec{g})$ for some vector $\hat{v}$, in which case the gauge field along $\hat{v}$ remains gapless. The number of independent vectors $\hat{v}$ satisfying this criterion is limited to be (at most) the dimension of the Cartan subalgebra $\mathcal{H}\subset \mathcal{G}$.

Assuming once more that a gauge symmetry remains, we can set $\hat{v}$ so $\hat{v}\shortdot \vec{g}\in \mathcal{H}$. Let $\hat{v}_i$ be a basis for the Cartan subalgebra. Then $\hat{v}_i\shortdot \vec{g}\ket{u}=r_i\ket{u}$ where $r_i$ is a weight of representation $m$ of Lie algebra $\mathcal{G}$. Then, after integrating out the gapped degrees of freedom we have
\begin{equation}
\label{eqn:Killing-form=K-matrix}
L_{eff} = - \frac{1}{\pi l} \epsilon^{\mu\nu i} A_\mu \partial_\nu \left(\vec{a}^H_i\shortdot r)\right)+\frac{k}{4\pi I_\text{ad}}\epsilon^{i0j}\vec{a}^H_i\shortdot K \shortdot \partial_0  \vec{a}^H_j
\end{equation}
where $K$ and $\vec{a}^H$ are the restrictions of, respectively, the Killing form and $\vec{a}$ to the Cartan subalgebra and $I_\text{ad}$ is the Dynkin index of the adjoint representation.
This Lagrangian must be supplemented by the Chern-Simons constraint:
\begin{equation}
  -\frac{\vec{r}}{\pi l}\epsilon^{0ij}\partial_i A_j+\frac{k}{4\pi  I_\text{ad}}\epsilon^{0ij} K \shortdot\partial_i \vec{a}^H_j=0
\end{equation}
that arises from eliminating gapped degrees of freedom in Eqs.~(\ref{eq-grestrict})~and~(\ref{eq-gflux}).
Incorporating this constraint into the Lagrangian through the use of a new Lagrange multiplier field $\vec{a}_0$ and redefining $\vec{a}_i=\vec{a}^H_i$ gives the condensed, gauge invariant Lagrange density
\begin{equation}
L_{eff} = - \frac{1}{2\pi} \epsilon^{\mu\nu \lambda}A_\mu \partial_\nu \left(\frac{2}{l}\vec{r}\shortdot\vec{a}_\lambda \right)+\frac{k}{4\pi I_\text{ad}}\epsilon^{\mu\nu\lambda}\vec{a}_\mu\shortdot K \cdot \partial_\nu  \vec{a}_\lambda
\end{equation}
The filling fraction of the resulting state is (for $l\neq0$)
\begin{equation}
\nu=\frac{4 I_\text{ad}}{l^2 k}\vec{r}\shortdot K^{-1}\shortdot\vec{r}
\end{equation}
Note that the factor of $4$ on the right-hand-side is due to the assumption that
the superconducting condensate has charge-$2$.
However, the resulting set of free quasiparticles may be distinct from that initially implied by the matrix $K$. We have thus far ignored the spatial variation of the internal state vector $\ket{u}$. In most cases, it is safe to do so because any such variation may be gauged away using a transformation in the original gauge group $G$. However, if there is a quasiparticle present that has non-trivial statistics with the condensing field $\Phi^m_l$,\footnote{That is, the squared $R$ matrix is not $\exp(2\pi i n)$ times the identity for n integer.} the resulting twist in $\ket{u}$ may not be gauged away by a non-singular gauge transformation. This results in an energy cost linearly divergent in the system size, resulting from the term $-|\nabla_i\Phi|^2$ in the original Lagrangian. (Recall that the magnitude of $\Phi$ is fixed by the condensation.)

\subsection{Example: $G_2$ level $1$}
\label{secG2}

As an example, we consider the case of a superconductor with
anyons governed by $G_2$ at level $1$. The group $G_2$
is $14$-dimensional, with a $2$-dimensional Cartan subalgebra.
This is a Fibonacci superconductor built on a bosonic quantum Hall state.
There are no fermions in the superconducting state;
the only non-trivial particle is a Fibonacci anyon, which carries the $7$-dimensional fundamental
representation of $G_2$. There are $14$ gauge fields in the
effective action (\ref{eqn:CS-LG-general-group}). When a composite of a Fibonacci
anyon and flux $l=1$ condenses, there are two remaining gapless gauge fields.
The $2\times 2$ K-matrix for these gapless gauge fields are obtained from the
Killing form, according to Eq. (\ref{eqn:Killing-form=K-matrix}). The Cartan matrix of
$G_2$ is:
\begin{equation}
\begin{pmatrix}
2 & -3 \\
-1 & 2
\end{pmatrix}
\end{equation}
After a rescaling of the bottom row by $3$, to remove the norms of the roots that enters the
expression for the Cartan matrix, we obtain
\begin{equation}
K =\begin{pmatrix}
2 & -3 \\
-3 & 6
\end{pmatrix}
\end{equation}
which is equal, after an $SL(2,\mathbb{Z})$ basis change,
to $K$-matrix:
\begin{equation}
K =\begin{pmatrix}
2 & 1 \\
1 & 2
\end{pmatrix}
\end{equation}
Under this basis change, the weight vector $\vec{r} = (1,-1)$ is transformed to the standard charge vector
for the $(2,2,1)$ state. \footnote{Other choices of weight vector can lead to a state with the same $K$ matrix and charge vector (1,2) for $r=(1, 0)$, giving integer filling. Alternately, the weight $r=(0,0)$ leads to an orbifold state with $U(1)\times U(1)\rtimes \mathbb{D}_3$ gauge symmetry. The final possibility, $r=(0,1)$, is equivalent to the weight used in the text under a $\mathbb{Z}_3$ symmetry of the $K$ matrix.}

In other words, we obtain the inverse of the construction in Ref. \onlinecite{Mong14a}.

\section{Discussion}

Recent progress has opened a promising route to non-Abelian topological phases: the liberation
of defects in Abelian topological phases. The coupled-chain construction\cite{Teo14,Mong14a,Sagi14,Stoudenmire15,Mross15} is a
concrete model for this scenario which has the virtue of solubility that is inherited from
one-dimensional theories. A Fibonacci supercoducting state can be constructed in this manner \cite{Mong14a}.
In this paper, we construct a manifestly two-dimensional Chern-Simons theory of this state.
As have seen in the preceeding sections, it is one member in a family of interesting topological phases
with corresponding $SU(2)_{p+q}\otimes SU(2)_{p-q}\vert_R $
Chern-Simons theories. Through Chern-Simons-Higgs transitions,
various superconducting and fractional quantum Hall states,
both Abelian and non-Abelian, are accessible through direct phase transitions.
The transitions are controlled by the type of quasiparticle that condenses
which, in our effective field theory, corresponds to the
gauge group representation carried by the condensing field.
We note that this description is dependent on the particular
Chern-Simons representation that we use. An example of this is given
by $SU(2)_4$, in which condensation of the particle carrying the $j=2$
representation of $SU(2)$ leads to an Abelian theory with remaining gauge
group $U(1)$ (coming from the Cartan subalgebra of $SU(2)$) and central charge
$c=1$. This is distinctly different from the Bais-Slingerland condensation scheme \cite{Bais09} on the same topological phase, which results in a $c=2$ theory. 
We can recover the Bais-Slingerland result
by representing the same anyon model by $U(1)\times U(1) \ltimes \mathbb{Z}_2$
\cite{Barkeshli10}. The $\mathbb{Z}_2$ boson in the metaplectic theory is related to the $j=2$ quasiparticle in the $SU(2)_4$ CS theory by the attachment of a non-Abelian gauge flux, in much the way that an electron is related to a `composite boson' in Landau-Ginzburg theories of the (Abelian) fractional quantum Hall effect \cite{Zhang89,Read89a}.

In the case of the Fibonacci superconductor, when a non-Abelian quasiparticle
carrying non-zero magnetic flux condenses, the system enters the $\nu=2/3$ Abelian
fractional quantum Hall state. Hence, this is the inverse of the transition from
embodied by the coupled-chain construction. This identification suggests
that our entire family of $SU(2)_{p+q}\otimes SU(2)_{p-q}\vert_R $ non-Abelian topological phases
is accessible by coupled chain constructions from Abelian $(p,p,q)$ fractional
quantum Hall states. The connection between our effective theories and coupled chain constructions
could be cemented by finding a direct correspondence between the condensing field
$\Phi^{(\frac{m}{2},\frac{n}{2})}_l$ and the 1D primary fields through which chains are coupled.
It would also be enlightening to construct the 2D effective theory dual to ours, in which
the particles in a $(p,p,q)$ Abelian topological phase fractionalize, leading to the
$SU(2)_{p+q}\otimes SU(2)_{p-q}\vert_R $ non-Abelian topological phase.
Such a theory could follow naturally from a parton construction, which would
also suggest trial wavefunctions for our non-Abelian phases.
At any rate, our work gives further impetus to the
search for semiconductor/superconductor hybrid systems in which fractional quantum Hall states
can be brought into contact with superconductivity.

\section{Acknowledgements}
D.J.C gratefully acknowledges the support of LPS-CMTC. We are grateful to Fiona Burnell, Michael Freedman, Kirill Shtengel, and Zhenghan Wang for helpful conversations.

\appendix
\section{Fermionic theories}\label{appfermions}
One way of describing fermionic anyon theories is as restrictions of bosonic theories. These restriction disallows all but a subset of the anyons of the bosonic theory that is closed under fusion and includes both a fermionic anyon and the identity sector. As a simple example of how such a restriction might arise, consider the $\nu=1$ quantum Hall state, with edge Lagrangian

\begin{equation}
L_{K=1}=\int d^2x \partial_x \phi (\partial_t\phi-v\partial_x\phi)
\end{equation}
Here $\phi$ is an angular variable compactified on $\left[0,2\pi\right)$, so that allowed operators include only derivatives of $\phi$ or the combinations $e^{i n\phi}$ for integer $n$.

If we rescale $\phi\rightarrow 2\phi$, then $\phi$ is now instead compactified on $\left[0,\pi\right)$, allowing only $e^{2i n\phi}$ for integer $n$. Meanwhile, the Lagrangian becomes
\begin{equation}
L_{K=4}=\int d^2x 4\partial_x \phi (\partial_t\phi-v\partial_x\phi),
\end{equation}
which, if $\phi$ had its original compactification, would be the Lagrangian of the $\mathbb{Z}_4^{(1/2)}$ theory of Table~\ref{Z4table}.

\begin{table}[htb]
{\setlength{\topsep}{-\parskip}
\setlength{\partopsep}{0pt}
\begin{tabularx}{.75\columnwidth}{|l|*{4}{|X}|}
\hline
\multicolumn{5}{|l|}{\hspace{4mm}$\mathbb{Z}_4^{(1/2)}$ \hspace{8mm} $c=1$}\\
\hline
  & \underline{$I$} & $a$  & \underline{$a^2$} & $a^3$ \\
\hline
$h$ & \underline{$0$} & $1/8$ & \underline{$1/2$} & $1/8$ \\
\hline\hline
\multicolumn{5}{|l|}{Fusion rules}\\
\hline
\multicolumn{5}{|p{.73\columnwidth}|}{
\begin{tabbing}
$a^n\otimes a^m=a^{m+n}$\quad $a^4=I$
\end{tabbing}}\\
\hline
\end{tabularx}}
\caption{The fields of the Abelian theory  $\mathbb{Z}_4^{(1/2)}$, with central charge $c=1$, along with their conformal spin $h$, and non-trivial fusion rules. Underlined fields form a closed set under fusion. We label this set as $\mathbb{Z}_4^{(1/2)}\vert_R$ and refer to it as the fermionic $\mathbb{Z}_4$ theory with central charge $1$}
\label{Z4table}
\end{table}
Note that if we look only at the restricted sector of the $\mathbb{Z}_4^{(1/2)}$ theory, there are only two fields, one of which is the identity. The other is $\psi\equiv a^2$, with spin $1/2$ (mod 1). The only non-trivial fusion rule is that
\begin{equation}
\psi\otimes\psi=I.
\end{equation}
The purpose of this Appendix is to catalogue the set of modular tensor categories that act as minimal modular extensions of the theory with one fermion, that is, theories into which the set $\{I,\psi\}$ can be embedded in such a way that fusion with any field outside of the set gives the whole (modular) theory. Any such theory can be restricted so that the fermion is the only `allowed' particle. By cataloguing these theories, we find the set of central charges that a single fermion can carry. There are sixteen such theories, with central charge \mbox{n/2 $\mathrm{mod}$ 8} are listed in Table~\ref{Fermiontables}. These are exactly the theories of Kitaev's 16-fold way \cite{Kitaev06a}.
\begin{table}[htb]
{\setlength{\topsep}{-\parskip}
\setlength{\partopsep}{0pt}
\begin{tabularx}{.75\columnwidth}{|l|*{4}{|X}|}
\hline
\multicolumn{4}{|l|}{\hspace{2mm}Ising-like theories\hspace{8mm}$c=n+1/2$}\\
\hline
  & \underline{$I$} & $\sigma$  & \underline{$\psi$} \\
\hline
$h$ & \underline{$0$} & $(2n+1)/16$ & \underline{$1/2$}\\
$d$ & \underline{$1$} & $\sqrt{2}$ & \underline{$1$}\\
\hline\hline
\multicolumn{4}{|l|}{Fusion rules}\\
\hline
\multicolumn{4}{|p{.73\columnwidth}|}{
\begin{tabbing}
$\sigma \otimes \sigma= I \oplus \psi$ \hspace{10mm}\=$\psi\otimes \sigma=\sigma$\\
$\psi\otimes \psi =I$
\end{tabbing}}\\
\hline
\end{tabularx}\vspace{.2in}
\begin{tabularx}{.75\columnwidth}{|l|*{4}{|X}|}
\hline
\multicolumn{5}{|l|}{\hspace{4mm}$\mathbb{Z}_4^{(n+1/2)}$ \hspace{8mm} $c=2n+1$}\\
\hline
  & \underline{$I$} & $a$  & \underline{$a^2$} & $a^3$ \\
\hline
$h$ & \underline{$0$} & $(2n+1)/8$ & \underline{$1/2$} & $(2n+1)/8$ \\
\hline\hline
\multicolumn{5}{|l|}{Fusion rules}\\
\hline
\multicolumn{5}{|p{.73\columnwidth}|}{
\begin{tabbing}
$a^n\otimes a^m=a^{m+n}$\quad $a^4=I$
\end{tabbing}}\\
\hline
\end{tabularx}\vspace{.2in}
\begin{tabularx}{.75\columnwidth}{|l|*{4}{|X}|}
\hline
\multicolumn{5}{|l|}{\hspace{4mm}$\mathbb{Z}_2^{(n+1/2)}\otimes\mathbb{Z}_2^{(n+1/2)}$ \hspace{8mm} $c=4n+2$}\\
\hline
  & \underline{$I$} & $a$  & \underline{$ab$} & $b$ \\
\hline
$h$ & \underline{$0$} & $(2n+1)/4$ & \underline{$1/2$} & $(2n+1)/4$ \\
\hline\hline
\multicolumn{5}{|l|}{Fusion rules}\\
\hline
\multicolumn{5}{|p{.73\columnwidth}|}{
\begin{tabbing}
$a \otimes a= I $ \hspace{10mm}\=$b \otimes b= I $\\
$a\otimes b =ab$\> $ab \otimes ab = I $
\end{tabbing}}\\
\hline
\end{tabularx}\vspace{.2in}
\begin{tabularx}{.75\columnwidth}{|l|*{4}{|X}|}
\hline
\multicolumn{5}{|l|}{\hspace{4mm}$D'(\mathbb{Z}_2)$ \hspace{8mm} $c=4$}\\
\hline
  & \underline{$I$} & $a$  & \underline{$ab$} & $b$ \\
\hline
$h$ & \underline{$0$} & $1/2$ & \underline{$1/2$} & $1/2$ \\
\hline\hline
\multicolumn{5}{|l|}{Fusion rules}\\
\hline
\multicolumn{5}{|p{.73\columnwidth}|}{
\begin{tabbing}
$a \otimes a= I $ \hspace{10mm}\=$b \otimes b= I $\\
$a\otimes b =ab$\> $ab \otimes ab = I $
\end{tabbing}}\\
\hline
\end{tabularx}\vspace{.2in}
\begin{tabularx}{.75\columnwidth}{|l|*{4}{|X}|}
\hline
\multicolumn{5}{|l|}{\hspace{4mm}$D(\mathbb{Z}_2)$ \hspace{8mm} $c=0$}\\
\hline
  & \underline{$I$} & $a$  & \underline{$ab$} & $b$ \\
\hline
$h$ & \underline{$0$} & $0$ & \underline{$1/2$} & $0$ \\
\hline\hline
\multicolumn{5}{|l|}{Fusion rules}\\
\hline
\multicolumn{5}{|p{.73\columnwidth}|}{
\begin{tabbing}
$a \otimes a= I $ \hspace{10mm}\=$b \otimes b= I $\\
$a\otimes b =ab$\> $ab \otimes ab = I $
\end{tabbing}}\\
\hline
\end{tabularx}}
\caption{The sixteen fusion theories at central charge \mbox{n/2 $\mathrm{mod} 8$}, for which the simple fermion is a maximal subtheory. Fields of the restricted (fermionic) theory are underlined. The eight Ising-like theories are non-Abelian, while the other eight theories are Abelian. The restricted theory is always Abelian and always has the same fusion rule $\psi\otimes\psi=I$.}
\label{Fermiontables}
\end{table}

\bibliography{topo-phases}

\begin{thebibliography}{29}
\expandafter\ifx\csname natexlab\endcsname\relax\def\natexlab#1{#1}\fi
\expandafter\ifx\csname bibnamefont\endcsname\relax
  \def\bibnamefont#1{#1}\fi
\expandafter\ifx\csname bibfnamefont\endcsname\relax
  \def\bibfnamefont#1{#1}\fi
\expandafter\ifx\csname citenamefont\endcsname\relax
  \def\citenamefont#1{#1}\fi
\expandafter\ifx\csname url\endcsname\relax
  \def\url#1{\texttt{#1}}\fi
\expandafter\ifx\csname urlprefix\endcsname\relax\def\urlprefix{URL }\fi
\providecommand{\bibinfo}[2]{#2}
\providecommand{\eprint}[2][]{\url{#2}}

\bibitem[{\citenamefont{{Bais} and {Slingerland}}(2009)}]{Bais09}
\bibinfo{author}{\bibfnamefont{F.~A.} \bibnamefont{{Bais}}} \bibnamefont{and}
  \bibinfo{author}{\bibfnamefont{J.~K.} \bibnamefont{{Slingerland}}},
  \bibinfo{journal}{\prb} \textbf{\bibinfo{volume}{79}}, \bibinfo{eid}{045316}
  (\bibinfo{year}{2009}), \eprint{0808.0627}.

\bibitem[{\citenamefont{Alicea}(2012)}]{Alicea12a}
\bibinfo{author}{\bibfnamefont{J.}~\bibnamefont{Alicea}},
  \bibinfo{journal}{Rep. Prog. Phys.} \textbf{\bibinfo{volume}{75}},
  \bibinfo{pages}{076501} (\bibinfo{year}{2012}), \eprint{arXiv:1202.1293}.

\bibitem[{\citenamefont{Beenakker}(2013)}]{Beenakker2013a}
\bibinfo{author}{\bibfnamefont{C.~W.~J.} \bibnamefont{Beenakker}},
  \bibinfo{journal}{Annu. Rev. Condens. Matter Phys.}
  \textbf{\bibinfo{volume}{4}}, \bibinfo{pages}{113} (\bibinfo{year}{2013}),
  \eprint{arXiv:1112.1950}.

\bibitem[{\citenamefont{{Das Sarma} et~al.}(2015)\citenamefont{{Das Sarma},
  {Freedman}, and {Nayak}}}]{DasSarma15}
\bibinfo{author}{\bibfnamefont{S.}~\bibnamefont{{Das Sarma}}},
  \bibinfo{author}{\bibfnamefont{M.}~\bibnamefont{{Freedman}}},
  \bibnamefont{and} \bibinfo{author}{\bibfnamefont{C.}~\bibnamefont{{Nayak}}},
  \bibinfo{journal}{ArXiv e-prints}  (\bibinfo{year}{2015}),
  \eprint{1501.02813}.

\bibitem[{\citenamefont{Teo and Kane}(2014)}]{Teo14}
\bibinfo{author}{\bibfnamefont{J.~C.~Y.} \bibnamefont{Teo}} \bibnamefont{and}
  \bibinfo{author}{\bibfnamefont{C.~L.} \bibnamefont{Kane}},
  \bibinfo{journal}{Phys. Rev. B} \textbf{\bibinfo{volume}{89}},
  \bibinfo{pages}{085101} (\bibinfo{year}{2014}).

\bibitem[{\citenamefont{{Mong} et~al.}(2014)\citenamefont{{Mong}, {Clarke},
  {Alicea}, {Lindner}, {Fendley}, {Nayak}, {Oreg}, {Stern}, {Berg}, {Shtengel}
  et~al.}}]{Mong14a}
\bibinfo{author}{\bibfnamefont{R.~S.~K.} \bibnamefont{{Mong}}},
  \bibinfo{author}{\bibfnamefont{D.~J.} \bibnamefont{{Clarke}}},
  \bibinfo{author}{\bibfnamefont{J.}~\bibnamefont{{Alicea}}},
  \bibinfo{author}{\bibfnamefont{N.~H.} \bibnamefont{{Lindner}}},
  \bibinfo{author}{\bibfnamefont{P.}~\bibnamefont{{Fendley}}},
  \bibinfo{author}{\bibfnamefont{C.}~\bibnamefont{{Nayak}}},
  \bibinfo{author}{\bibfnamefont{Y.}~\bibnamefont{{Oreg}}},
  \bibinfo{author}{\bibfnamefont{A.}~\bibnamefont{{Stern}}},
  \bibinfo{author}{\bibfnamefont{E.}~\bibnamefont{{Berg}}},
  \bibinfo{author}{\bibfnamefont{K.}~\bibnamefont{{Shtengel}}},
  \bibnamefont{et~al.}, \bibinfo{journal}{Physical Review X}
  \textbf{\bibinfo{volume}{4}}, \bibinfo{eid}{011036} (\bibinfo{year}{2014}),
  \eprint{1307.4403}.

\bibitem[{\citenamefont{Haldane}(1983)}]{Haldane83}
\bibinfo{author}{\bibfnamefont{F.~D.~M.} \bibnamefont{Haldane}},
  \bibinfo{journal}{Phys. Rev. Lett.} \textbf{\bibinfo{volume}{51}},
  \bibinfo{pages}{605} (\bibinfo{year}{1983}).

\bibitem[{\citenamefont{Halperin}(1984)}]{Halperin84}
\bibinfo{author}{\bibfnamefont{B.~I.} \bibnamefont{Halperin}},
  \bibinfo{journal}{Phys. Rev. Lett.} \textbf{\bibinfo{volume}{52}},
  \bibinfo{pages}{1583} (\bibinfo{year}{1984}).

\bibitem[{\citenamefont{{Jain}}(2007)}]{Jain07}
\bibinfo{author}{\bibfnamefont{J.~K.} \bibnamefont{{Jain}}},
  \emph{\bibinfo{title}{{Composite Fermions}}} (\bibinfo{publisher}{Cambridge
  University Press}, \bibinfo{address}{Cambridge, UK}, \bibinfo{year}{2007}).

\bibitem[{\citenamefont{Bonderson and Slingerland}(2008)}]{Bonderson08a}
\bibinfo{author}{\bibfnamefont{P.}~\bibnamefont{Bonderson}} \bibnamefont{and}
  \bibinfo{author}{\bibfnamefont{J.~K.} \bibnamefont{Slingerland}},
  \bibinfo{journal}{Phys. Rev. B} \textbf{\bibinfo{volume}{78}},
  \bibinfo{pages}{125323} (\bibinfo{year}{2008}), \eprint{arXiv:0711.3204}.

\bibitem[{\citenamefont{{Vaezi} and {Barkeshli}}(2014)}]{Vaezi14}
\bibinfo{author}{\bibfnamefont{A.}~\bibnamefont{{Vaezi}}} \bibnamefont{and}
  \bibinfo{author}{\bibfnamefont{M.}~\bibnamefont{{Barkeshli}}},
  \bibinfo{journal}{Physical Review Letters} \textbf{\bibinfo{volume}{113}},
  \bibinfo{eid}{236804} (\bibinfo{year}{2014}), \eprint{1403.3383}.

\bibitem[{\citenamefont{{Sagi} and {Oreg}}(2014)}]{Sagi14}
\bibinfo{author}{\bibfnamefont{E.}~\bibnamefont{{Sagi}}} \bibnamefont{and}
  \bibinfo{author}{\bibfnamefont{Y.}~\bibnamefont{{Oreg}}},
  \bibinfo{journal}{\prb} \textbf{\bibinfo{volume}{90}}, \bibinfo{eid}{201102}
  (\bibinfo{year}{2014}), \eprint{1403.1791}.

\bibitem[{\citenamefont{{Stoudenmire} et~al.}(2015)\citenamefont{{Stoudenmire},
  {Clarke}, {Mong}, and {Alicea}}}]{Stoudenmire15}
\bibinfo{author}{\bibfnamefont{E.~M.} \bibnamefont{{Stoudenmire}}},
  \bibinfo{author}{\bibfnamefont{D.~J.} \bibnamefont{{Clarke}}},
  \bibinfo{author}{\bibfnamefont{R.~S.~K.} \bibnamefont{{Mong}}},
  \bibnamefont{and} \bibinfo{author}{\bibfnamefont{J.}~\bibnamefont{{Alicea}}},
  \bibinfo{journal}{ArXiv e-prints}  (\bibinfo{year}{2015}),
  \eprint{1501.05305}.

\bibitem[{\citenamefont{{Mross} et~al.}(2015)\citenamefont{{Mross}, {Essin},
  and {Alicea}}}]{Mross15}
\bibinfo{author}{\bibfnamefont{D.~F.} \bibnamefont{{Mross}}},
  \bibinfo{author}{\bibfnamefont{A.}~\bibnamefont{{Essin}}}, \bibnamefont{and}
  \bibinfo{author}{\bibfnamefont{J.}~\bibnamefont{{Alicea}}},
  \bibinfo{journal}{Physical Review X} \textbf{\bibinfo{volume}{5}},
  \bibinfo{eid}{011011} (\bibinfo{year}{2015}), \eprint{1410.4201}.

\bibitem[{\citenamefont{Fradkin et~al.}(1999)\citenamefont{Fradkin, Nayak, and
  Schoutens}}]{Fradkin99b}
\bibinfo{author}{\bibfnamefont{E.}~\bibnamefont{Fradkin}},
  \bibinfo{author}{\bibfnamefont{C.}~\bibnamefont{Nayak}}, \bibnamefont{and}
  \bibinfo{author}{\bibfnamefont{K.}~\bibnamefont{Schoutens}},
  \bibinfo{journal}{Nucl. Phys. B} \textbf{\bibinfo{volume}{546}},
  \bibinfo{pages}{711} (\bibinfo{year}{1999}),
  \bibinfo{note}{cond-mat/9811005}.

\bibitem[{\citenamefont{Read and Green}(2000)}]{Read00}
\bibinfo{author}{\bibfnamefont{N.}~\bibnamefont{Read}} \bibnamefont{and}
  \bibinfo{author}{\bibfnamefont{D.}~\bibnamefont{Green}},
  \bibinfo{journal}{Phys. Rev. B} \textbf{\bibinfo{volume}{61}},
  \bibinfo{pages}{10267} (\bibinfo{year}{2000}).

\bibitem[{\citenamefont{{Barkeshli} et~al.}()\citenamefont{{Barkeshli},
  {Bonderson}, {Cheng}, and {Wang}}}]{Barkeshli14}
\bibinfo{author}{\bibfnamefont{M.}~\bibnamefont{{Barkeshli}}},
  \bibinfo{author}{\bibfnamefont{P.}~\bibnamefont{{Bonderson}}},
  \bibinfo{author}{\bibfnamefont{M.}~\bibnamefont{{Cheng}}}, \bibnamefont{and}
  \bibinfo{author}{\bibfnamefont{Z.}~\bibnamefont{{Wang}}},
  \emph{\bibinfo{title}{{Symmetry, Defects, and Gauging of Topological
  Phases}}}, \bibinfo{note}{arXiv:1410.4540}.

\bibitem[{\citenamefont{{Barkeshli} and {Wen}}(2010)}]{Barkeshli10}
\bibinfo{author}{\bibfnamefont{M.}~\bibnamefont{{Barkeshli}}} \bibnamefont{and}
  \bibinfo{author}{\bibfnamefont{X.-G.} \bibnamefont{{Wen}}},
  \bibinfo{journal}{\prb} \textbf{\bibinfo{volume}{81}}, \bibinfo{eid}{045323}
  (\bibinfo{year}{2010}), \eprint{0909.4882}.

\bibitem[{\citenamefont{Rowell and Wang}(2012)}]{Rowell12}
\bibinfo{author}{\bibfnamefont{E.~C.} \bibnamefont{Rowell}} \bibnamefont{and}
  \bibinfo{author}{\bibfnamefont{Z.}~\bibnamefont{Wang}},
  \bibinfo{journal}{Comm. Math. Phys.} \textbf{\bibinfo{volume}{311}},
  \bibinfo{pages}{595} (\bibinfo{year}{2012}).

\bibitem[{\citenamefont{Naidu and Rowell}(2011)}]{Naidu11}
\bibinfo{author}{\bibfnamefont{D.}~\bibnamefont{Naidu}} \bibnamefont{and}
  \bibinfo{author}{\bibfnamefont{E.~C.} \bibnamefont{Rowell}},
  \bibinfo{journal}{Algebr. Represent. Theory} \textbf{\bibinfo{volume}{15}},
  \bibinfo{pages}{837} (\bibinfo{year}{2011}).

\bibitem[{\citenamefont{Hastings et~al.}(2013)\citenamefont{Hastings, Nayak,
  and Wang}}]{Hastings13}
\bibinfo{author}{\bibfnamefont{M.~B.} \bibnamefont{Hastings}},
  \bibinfo{author}{\bibfnamefont{C.}~\bibnamefont{Nayak}}, \bibnamefont{and}
  \bibinfo{author}{\bibfnamefont{Z.}~\bibnamefont{Wang}},
  \bibinfo{journal}{Phys. Rev. B} \textbf{\bibinfo{volume}{87}},
  \bibinfo{pages}{165421} (\bibinfo{year}{2013}), \eprint{arXiv:1210.5477}.

\bibitem[{\citenamefont{Clarke et~al.}(2013)\citenamefont{Clarke, Alicea, and
  Shtengel}}]{Clarke13a}
\bibinfo{author}{\bibfnamefont{D.~J.} \bibnamefont{Clarke}},
  \bibinfo{author}{\bibfnamefont{J.}~\bibnamefont{Alicea}}, \bibnamefont{and}
  \bibinfo{author}{\bibfnamefont{K.}~\bibnamefont{Shtengel}},
  \bibinfo{journal}{Nat. Commun.} \textbf{\bibinfo{volume}{4}},
  \bibinfo{pages}{1348} (\bibinfo{year}{2013}), \eprint{arXiv:1204.5479}.

\bibitem[{\citenamefont{Lindner et~al.}(2012)\citenamefont{Lindner, Berg,
  Refael, and Stern}}]{Lindner12}
\bibinfo{author}{\bibfnamefont{N.~H.} \bibnamefont{Lindner}},
  \bibinfo{author}{\bibfnamefont{E.}~\bibnamefont{Berg}},
  \bibinfo{author}{\bibfnamefont{G.}~\bibnamefont{Refael}}, \bibnamefont{and}
  \bibinfo{author}{\bibfnamefont{A.}~\bibnamefont{Stern}},
  \bibinfo{journal}{Phys. Rev. X} \textbf{\bibinfo{volume}{2}},
  \bibinfo{pages}{041002} (\bibinfo{year}{2012}), \eprint{arXiv:1204.5733}.

\bibitem[{\citenamefont{Cheng}(2012)}]{Cheng12}
\bibinfo{author}{\bibfnamefont{M.}~\bibnamefont{Cheng}},
  \bibinfo{journal}{Phys. Rev. B} \textbf{\bibinfo{volume}{86}},
  \bibinfo{pages}{195126} (\bibinfo{year}{2012}), \eprint{arXiv:1204.6084}.

\bibitem[{\citenamefont{{Cui} and {Wang}}(2015)}]{Cui15}
\bibinfo{author}{\bibfnamefont{S.~X.} \bibnamefont{{Cui}}} \bibnamefont{and}
  \bibinfo{author}{\bibfnamefont{Z.}~\bibnamefont{{Wang}}},
  \bibinfo{journal}{Journal of Mathematical Physics}
  \textbf{\bibinfo{volume}{56}}, \bibinfo{eid}{032202} (\bibinfo{year}{2015}),
  \eprint{1405.7778}.

\bibitem[{\citenamefont{{Levaillant} et~al.}()\citenamefont{{Levaillant},
  {Bauer}, {Freedman}, {Wang}, and {Bonderson}}}]{Levaillant15}
\bibinfo{author}{\bibfnamefont{C.}~\bibnamefont{{Levaillant}}},
  \bibinfo{author}{\bibfnamefont{B.}~\bibnamefont{{Bauer}}},
  \bibinfo{author}{\bibfnamefont{M.}~\bibnamefont{{Freedman}}},
  \bibinfo{author}{\bibfnamefont{Z.}~\bibnamefont{{Wang}}}, \bibnamefont{and}
  \bibinfo{author}{\bibfnamefont{P.}~\bibnamefont{{Bonderson}}},
  \emph{\bibinfo{title}{Fusion and measurement operations for $su(2)_4$
  anyons}}, \bibinfo{note}{arXiv:1504.02098}.

\bibitem[{\citenamefont{Zhang et~al.}(1989)\citenamefont{Zhang, Hansson, and
  Kivelson}}]{Zhang89}
\bibinfo{author}{\bibfnamefont{S.~C.} \bibnamefont{Zhang}},
  \bibinfo{author}{\bibfnamefont{T.~H.} \bibnamefont{Hansson}},
  \bibnamefont{and} \bibinfo{author}{\bibfnamefont{S.}~\bibnamefont{Kivelson}},
  \bibinfo{journal}{Phys. Rev. Lett.} \textbf{\bibinfo{volume}{62}},
  \bibinfo{pages}{82} (\bibinfo{year}{1989}).

\bibitem[{\citenamefont{Read}(1989)}]{Read89a}
\bibinfo{author}{\bibfnamefont{N.}~\bibnamefont{Read}}, \bibinfo{journal}{Phys.
  Rev. Lett.} \textbf{\bibinfo{volume}{62}}, \bibinfo{pages}{86}
  (\bibinfo{year}{1989}).

\bibitem[{\citenamefont{Kitaev}(2006)}]{Kitaev06a}
\bibinfo{author}{\bibfnamefont{A.~Y.} \bibnamefont{Kitaev}},
  \bibinfo{journal}{Ann. Phys. (N.Y.)} \textbf{\bibinfo{volume}{321}},
  \bibinfo{pages}{2} (\bibinfo{year}{2006}), \bibinfo{note}{cond-mat/0506438}.

\end{thebibliography}

\end{document}